\documentclass[12pt]{iopart}
\usepackage{graphicx}
\usepackage{url}
\begin{document}

\title{Spin-phonon coupling in antiferromagnetic chromium spinels}

\author{T Rudolf$^1$, Ch Kant$^1$, F Mayr$^1$, J Hemberger$^1$, V Tsurkan$^{1,2}$, and A Loidl$^1$}

\address{$^1$ Experimental Physics V, Center for Electronic Correlations and Magnetism, University of Augsburg, D-86135 Augsburg, Germany}
\address{$^2$ Institute of Applied Physics, Academy of Sciences of Moldova, MD-2028 Chisinau, Republic of Moldova}
\ead{torsten.rudolf@physik.uni-augsburg.de}

\begin{abstract}
The temperature dependence of eigenfrequencies and intensities of the IR active
modes has been investigated for the antiferromagnetic chromium spinel compounds
CdCr$_2$O$_4$, ZnCr$_2$O$_4$, ZnCr$_2$S$_4$, ZnCr$_2$Se$_4$, and HgCr$_2$S$_4$
by IR spectroscopy for temperatures from 5~K to 300~K. At the transition into
the magnetically ordered phases, and driven by spin-phonon coupling, most
compounds reveal significant splittings of the phonon modes. This is true for
geometrically frustrated CdCr$_2$O$_4$, and ZnCr$_2$O$_4$, for bond frustrated
ZnCr$_2$S$_4$ and for ZnCr$_2$Se$_4$, which also is bond frustrated, but
dominated by ferromagnetic exchange. The pattern of splitting is different for
the different compounds and crucially depends on the nature of frustration and
of the resulting spin order. HgCr$_2$S$_4$, which is almost ferromagnetic,
exhibits no splitting of the eigenfrequencies, but shows significant shifts due
to ferromagnetic spin fluctuations.
\end{abstract}

\maketitle

\section{Introduction}
It is known since more than 40 years that chromium spinels ($A$Cr$_2$X$_4$)
span an enormous range of magnetic exchange strengths and different magnetic
ground states~\cite{baltzer}. As a function of lattice constant, or
equivalently as a function of Cr-Cr separation, these compounds are
characterized by Curie-Weiss temperatures from -400~K to 200~K and, at low
temperatures, reveal either complex antiferromagnetism or ferromagnetism. The
chromium oxide-spinels undergo antiferromagnetic (AFM) ordering transitions of
order 10~K, despite the fact that the exchange interactions, as deduced from
the paramagnetic Curie-Weiss (CW) temperatures, are one order of magnitude
larger. This can be explained by the fact that the Cr spins reside on a
pyrochlore lattice revealing strong geometrical frustration. Some selenide
spinels undergo ferromagnetic (FM) ordering at temperatures of the order of
100~K. The different ground states as a function of lattice constants are
driven by the dominating exchange interactions: At small Cr-Cr separation
strong direct AFM exchange dominates. With increasing separation the 90$^\circ$
FM Cr-$X$-Cr exchange becomes important. Probably for all lattice spacings a
complex Cr-$X$-$A$-$X$-Cr super exchange (SE) is active. This antiferromagnetic
SE is weak and only of the order of 1~K, but gains importance via a high
multiplicity \cite{baltzer}.\

In the spinel structure, Cr ions are located at the B-sites in an octahedral
environment in a $3d^3$ state. Under the action of an octahedral crystal field
the $d$ levels split into a lower $t_{2g}$ triplet, each orbital being singly
occupied, and an excited empty $e_g$ doublet. The resulting bulk material is a
Mott insulator with $S=3/2$, with an almost spherical charge distribution and
negligible spin-orbit coupling. As a matter of fact, one can expect that in
these compounds orbital effects are quenched, and that charge effects will play
no role. The geometric frustration of the pyrochlore lattice and the bond
frustration due to competing magnetic exchange~\cite{hemberger2002} are
responsible for emergent complex magnetic ground states~\cite{lee}, for strong
metamagnetism~\cite{hemberger2006, tsurkan}, and even for multiferroic
behaviour~\cite{hemberger2005, weber}. The complexity of the ground states
becomes even enhanced by the unconventional coupling of the spin-only chromium
moments to the lattice. This has early been recognized on the basis of
ultrasonic experiments on ZnCr$_2$O$_4$ by Kino and L\"{u}thi~\cite{kino} and later
on, has been treated in modern theories on the basis of a spin-driven
Jahn-Teller effect~\cite{yamashita, tchernyshyov}.\

Very recently, spin-phonon coupling in materials with strong electronic
correlations gained considerable attention. Concerning the compounds under
consideration, the onset of structural distortions and of AFM order in the
geometrically frustrated oxides has been described using a concept, in which
the lattice distorts to gain exchange energy and thereby relieves the
frustration~\cite{yamashita, tchernyshyov, tchernyshyov66}. In ZnCr$_2$S$_4$
complex magnetic order results from strong bond-frustration characterised by
ferromagnetic FM and AFM exchange interactions of almost equal
strength~\cite{hemberger2002}. ZnCr$_2$Se$_4$ is dominated by FM exchange but
orders antiferromagnetically at $T_\textrm{N} = 21$~K~\cite{rudolf}. All these
compounds show a strong splitting of specific phonon modes~\cite{hemberger2002,
rudolf, sushkov}, which has been explained by ab-initio approaches providing
evidence that the dynamic anisotropy of the phonon modes is induced by the
magnetic exchange interactions alone and can be fully decoupled from lattice
distortions~\cite{massida, fennie}. The idea that magnetic exchange
interactions and the onset of magnetic order strongly influence the phonon
modes has been outlined already decades ago by Baltensperger and
Helman~\cite{baltensperger1968, baltensperger1970} and by Br\"{u}esch and
D'Ambrogio~\cite{bruesch}.\
\begin{table}
\caption{\label{tab1}Lattice constants $a_0$({\AA}), fractional coordinate $x$,
effective paramagnetic moment $p_{eff}(\mu_\textrm{B})$, Curie-Weiss
temperature $\Theta_\textrm{CW}$(K), magnetic ordering temperature $T_m$(K) and
magnetic frustration parameter $f=|\Theta_\textrm{CW}|/T_m$. The values
documented in this table have, if not otherwise referenced, been taken from our
own work, which mostly has been published~\cite{hemberger2002},
\cite{hemberger2006}-\cite{weber}.}
\begin{indented}
\lineup
\item[]\begin{tabular}{llllllll} \br
\centre{2}{Compound} & \centre{1}{$a_0$} & \centre{1}{$x$} & \centre{1}{$p_{eff}$} & \centre{1}{$\Theta_\textrm{CW}$} & \centre{1}{$T_m$} & \centre{1}{$f$}\\
\mr
ZnCr$_2$O$_4$ & ZCO & \08.317(2) & 0.265 & 3.85 & \-398 & 12.5 & 32\\
MgCr$_2$O$_4$ & MCO & \08.319(3) & 0.261 & 3.71 & \-346 & 12.7 & 27\\
CdCr$_2$O$_4$ & CCO & \08.596(2) & 0.265 & 4.03 & \0\-71 & \08.2 & 8.7\\
HgCr$_2$O$_4$ & HCO & \08.658(1)$^a$ & 0.229$^a$ & 3.72$^b$ & \0\-32$^b$ & \05.8$^b$ & 5.5\\
ZnCr$_2$S$_4$ & ZCS & \09.983(2) & 0.258 & 3.86 & 7.9 & 15; 8 & 0.5\\
ZnCr$_2$Se$_4$ & ZCSe & 10.498(2) & 0.260 & 4.04 & \090 & \021 & 4.3\\
HgCr$_2$S$_4$ & HCS & 10.256(1) & 0.267 & 3.90 & 140 & \022 & 6.4\\
CdCr$_2$S$_4$ & CCS & 10.247(2) & 0.263 & 3.88 & 155 & \075 & 2.1\\
CdCr$_2$Se$_4$ & CCSe & 10.740(3) & 0.264 & 3.82 & 184 & 130 & 1.4\\
HgCr$_2$Se$_4$ & HCSe & 10.737(3) & 0.264 & 3.89 & 200 & 106 & 1.9\\
\br
\end{tabular}\\
\noindent $^a$ Wessels {\it et al.}~\cite{wessels}\\
\noindent $^b$ Ueda {\it et al.}~\cite{ueda}\\
\end{indented}
\end{table}

In this Letter we report on detailed infrared (IR) experiments for a variety of
spinels, ranging from compounds with strong geometric frustration to almost
ferromagnetic systems. We document spin-phonon coupling by splittings and
significant shifts of different phonon modes, which are diverse for different
spin orders. We try to correlate the shifts of the eigenfrequencies and the
dynamic breaking of symmetry by the strength of competing exchange
interactions. Table~1 shows room temperature lattice constants and fractional
coordinates, paramagnetic moments, Curie-Weiss temperatures, magnetic ordering
temperatures, and the magnetic frustration parameter
$f=|\Theta_\textrm{CW}|/T_m$ for a number of chromium oxides, sulphides, and
selenides. From this table it immediately becomes clear, that with increasing
lattice constant the CW temperatures increase from strongly negative AFM
exchange in the oxides to moderately FM exchange in the selenides. As outlined
above, in the oxides AFM direct Cr-Cr exchange is dominating all other exchange
paths. With increasing lattice constant this direct exchange becomes
considerably weaker and competes with 90$^\circ$ FM Cr-$X$-Cr exchange. Hg and
Cd selenides are well known ferromagnetic semiconductors with high FM ordering
temperatures. At the same time, table~1 demonstrates that the chromium spins in
the oxides exhibit strong geometrical frustration. In all compounds, the Cr
moments are close to the spin-only values ($S=3/2$, yielding 3.87
$\mu_\textrm{B}$ for the paramagnetic moment) demonstrating the absence of
spin-orbit coupling.\

The evolution of the magnetic properties of the chromium spinels is
schematically shown in figure~1. Here we plot the characteristic ordering
temperatures vs. the Curie-Weiss temperatures. The CW temperatures scale with
the lattice constants (see table~1), but, for presentation purposes we decided
to take the CW temperatures as variables. In the complete regime from strongly
negative CW temperatures (-400~K) to positive CW temperatures of almost 150~K,
the ground states of all spinels exhibit antiferromagnetism with different spin
arrangements: ZCO has a complex magnetic order with at least 64 spins per
magnetic unit cell~\cite{lee, chern}. CCO undergoes a magnetic phase transition
into a phase with incommensurate spiral order, where a collinear state is
twisted into a long spiral with the spins aligned in the $ac$
plane~\cite{chern, chung}. In ZCS, FM and AFM exchange are almost of equal
strength~\cite{hemberger2002}. Consequently, the magnetic ground state is a
mixed state, with the coexistence of a spiral and a collinear
structure~\cite{hamedoun}. This mixed phase is established at 8~K. A magnetic
precursor phase with a pure helical spin order appears at 15~K~\cite{hamedoun}.
ZCSe and HCS are close to the FM border with CW temperatures of the order of
100~K. Both compounds are dominated by strong ferromagnetic spin fluctuations
but finally undergo AFM phase transitions with FM planes with a small turn
angle of the spin directions of neighbouring
planes~\cite{plumier1965}-\cite{chapon}. Finally CCS, CCSe, and HCSe are
ferromagnetic semiconductors with high ordering temperatures~\cite{baltzer}.\

\begin{figure}
\begin{center}
\includegraphics[width=13cm]{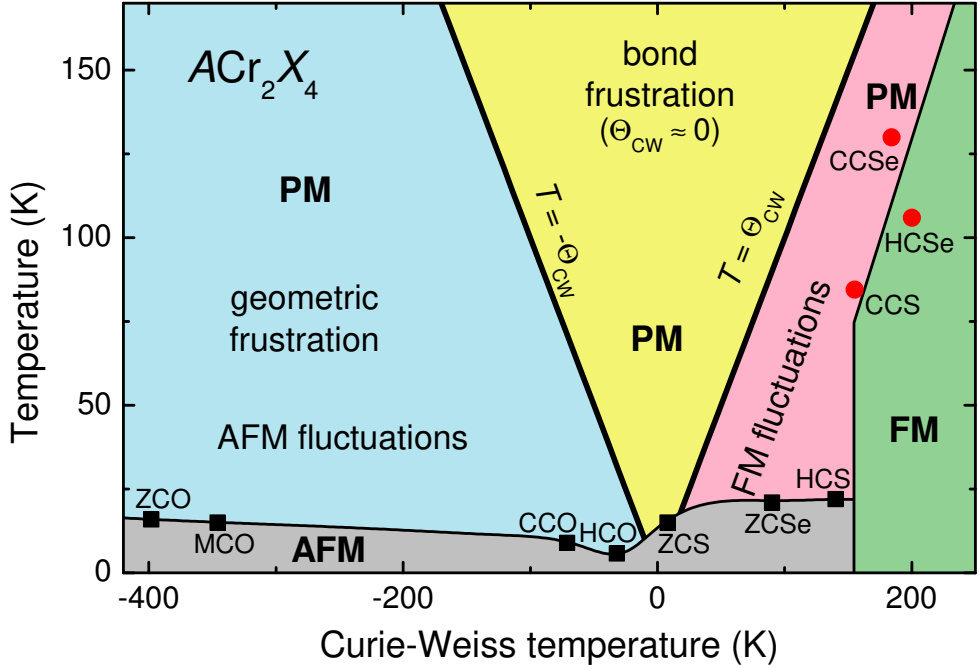}
\end{center}
\caption{\label{fig1}Schematic magnetic phase diagram of $A$Cr$_2X_4$
compounds, where characteristic temperatures are plotted vs. the Curie-Weiss
temperature (see table~\ref{tab1}): FM (red circles) and AFM (black squares)
ordering temperatures; hypothetical magnetic ordering temperatures
($T=\pm\Theta_\textrm{CW}$) are indicated by thick solid lines. Thin solid
lines separate magnetically ordered from paramagnetic phases and are drawn to
guide the eye.}
\end{figure}

The oxides are dominated by strong direct AFM exchange. As the spins reside on
a pyrochlore lattice, geometrical frustration suppresses conventional AFM
order. The FM compounds are dominated by 90$^\circ$ nearest neighbour (nn)
Cr-$X$-Cr exchange. In figure~1, the hypothetical FM and AFM ordering
temperatures corresponding to the Curie-Weiss temperatures are indicated by
thick solid lines. The temperature difference between this line and the real
ordering temperature is a direct measure of the frustration of the system. The
onset of AFM order is strongly reduced by geometrical frustration and in the
low-temperature paramagnetic phase AFM fluctuations dominate the physics of
theses compounds. FM order becomes suppressed by the onset of competing AFM
interactions and below the hypothetical ordering temperature FM fluctuations
control the low-temperature physics. In between these two lines with $T =
-\Theta_\textrm{CW}$ and $T = \Theta_\textrm{CW}$, the paramagnetic (PM) regime
extends to low temperatures because of bond frustration: Here FM exchange and
AFM Cr-$X$-$A$-$X$-Cr exchange, enhanced by residues of direct exchange, are of
equal strength. It is important to note that even for a vanishing CW
temperature local exchange interactions are still strong but compete and almost
cancel out each other. In the cone between these two lines of expected magnetic
order strong bond frustration is active, exemplified by the paramagnetic
behaviour of ZCS, which reveals a CW temperature of almost 0~K and a mixed
magnetic state at low temperatures~\cite{hamedoun}. Bond frustration still
seems to be active for compounds with high FM Curie-Weiss temperatures, which
exhibit antiferromagnetism at low temperatures. Finally we would like to
notice, that the spiral ground state in ZCSe has been explained assuming a set
of five distinct distant-neighbour interactions~\cite{dwight}.\

A closer inspection of table~1 and figure~1 allows further deep insights into
the dominating exchange paths: It is clear that the antiferromagnetic exchange
at small lattice constants, which is due to direct Cr-Cr overlap, becomes
exponentially smaller with increasing lattice spacings. However, it remains
unclear why the ferromagnetic exchange strongly increases towards larger
lattice constants. Table~1 shows that the lattice constants increase from
1.050~nm to 1.074~nm, when moving from ZCSe towards CCSe and HCSe, but the CW
temperatures increase from 90~K to 200~K and the ferromagnetic ordering
temperatures raise from 21~K to 130~K. The low ordering temperature for the
ZCSe can partly be explained by bond frustration, but it is hard to believe
that antiferromagnetic order via Cr-$X$-$A$-$X$-Cr depends so strongly on
distance. The ferromagnetic exchange should mainly depend on the Cr-$X$-Cr bond
angle, which is 90$^\circ$ in the ideal spinel structure. This bond angle
strongly depends on the fractional coordinate $x$~\cite{baltzer}. From table~1
we conclude that, as $x$ is almost constant for the ferromagnets under
consideration, it can not explain the difference in the FM exchange
interactions, and one can speculate that the increase in ferromagnetic exchange
rather comes from an increasing covalency of the Cr-$X$-Cr bonds when moving
from the sulfides to the selenides.\

\section{Experimental}
Polycrystalline CdCr$_2$O$_4$, ZnCr$_2$O$_4$, ZnCr$_2$S$_4$, ZnCr$_2$Se$_4$,
HgCr$_2$S$_4$, and CdCr$_2$S$_4$ samples were prepared by solid-state reaction
from the high purity elements in evacuated quartz ampoules. The synthesis was
repeated several times in order to reach good homogeneity.  For the sulfides
and selenides also single crystals have been grown by gas transport using
chlorine as transport agent or by liquid-transport methods. But only the size
of the CCS and ZCSe samples was large enough to perform high precision optical
experiments even at low wave numbers. X-ray diffraction revealed cubic
single-phase materials with lattice constants and anion fractional coordinates
as indicated in table~1. Magnetic susceptibilities were measured with a
commercial superconducting quantum interference device magnetometer for
temperatures from 1.8~K to 400~K. Curie-Weiss temperatures and paramagnetic
moments are included in table~1. Reflectivity experiments were carried out in
the far-infrared range using the Bruker Fourier-transform spectrometer IFS113v
equipped with a He bath cryostat. With our setup of mirrors and detectors we
were able to measure the frequency range from 50 to 650~cm$^{-1}$. The ceramic
samples were pressed with a maximal pressure of 1~GPa. Nevertheless, neither
surface nor the density of the ceramics was ideal and of perfect optical
quality. To correct for these factors we multiplied the measured reflectivities
of the ceramic samples CCO, ZCO, ZCS and HCS by a factor of 1.2. This factor
has been derived from a set of measurements of different sulfide- and
selenide-spinels at room temperature in ceramic and single-crystalline form.
The measurements on pressed ceramics hence limits the precision of the value of
$\epsilon_{\infty}$ and concomitantly of the dielectric strength. But the
absolute errors will be certainly less than 10\% and will not affect the
variation of the derived parameters as a function of temperature and external
magnetic field.\

\section{Results and discussion}
Reflectivity spectra as a function of temperature have been measured for all
six compounds with antiferromagnetic ground states. As expected from symmetry
grounds, the spectra consist of four phonon triplets of F$_{1u}$ symmetry.
Figures~2(a)-2(d) show the reflectivity spectra of CdCr$_2$O$_4$,
ZnCr$_2$S$_4$, ZnCr$_2$Se$_4$, and HgCr$_2$S$_4$ at room temperature, which
agree well with those previously measured~\cite{lutz, zwinscher}. The small
spikes in the reflectivity of CdCr$_2$O$_4$, which appear close to 440, 470 and
530~cm$^{-1}$, result from the surface roughness of the ceramic sample and show
no temperature dependence. In the following, the four modes are labeled as
mode~1 to 4 in ascending order from low to high wave numbers.\

\begin{figure}
\begin{center}
\includegraphics[width=10cm]{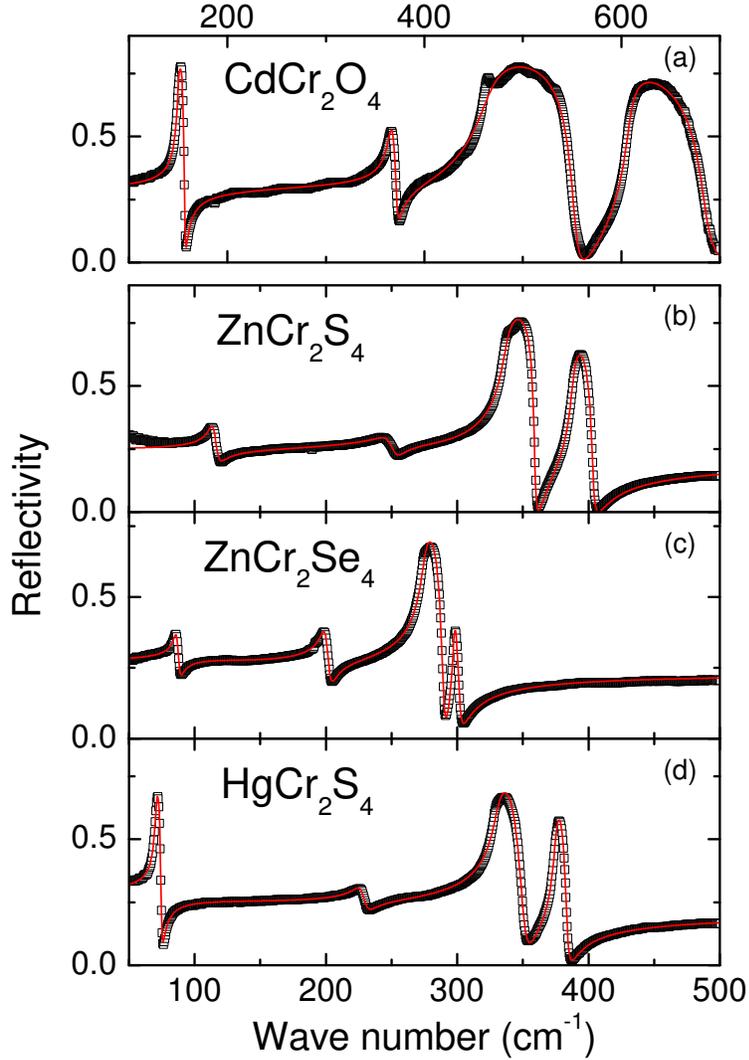}
\end{center}
\caption{\label{fig2}Room temperature reflectivity vs. wave number for
CdCr$_2$O$_4$ (a), ZnCr$_2$S$_4$ (b), ZnCr$_2$Se$_4$ (c), and HgCr$_2$S$_4$
(d). The solid lines represent the results of fits as described in the text.
Please note the change of the wave-number scale for frame (a).}
\end{figure}

To analyze the reflectivity spectra the complex dielectric function
$\epsilon(\omega)$ was obtained by calculating the factorized form
\begin{equation}
\label{equ1}
\epsilon(\omega)=\epsilon_{\infty}\prod_j\frac{\omega_{Lj}^2-\omega^2-\rmi\gamma_{Lj}\omega}{\omega_{Tj}^2-\omega^2-\rmi\gamma_{Tj}\omega}.
\end{equation}
Here $\omega_{Lj}$, $\omega_{Tj}$, $\gamma_{Lj}$ and $\gamma_{Tj}$
correspond to longitudinal ($L$) and transversal ($T$)
eigenfrequency ($\omega_j$) and damping ($\gamma_j$) of mode $j$,
respectively. At normal incidence the dielectric function is related
to the reflectivity via
\begin{equation}
\label{equ2}
R(\omega)=\left|\frac{\sqrt{\epsilon(\omega)}-1}{\sqrt{\epsilon(\omega)}+1}\right|^2.
\end{equation}
From the measured reflectivity the values of $\epsilon_{\infty}$,
$\omega_{Lj}$, $\omega_{Tj}$, $\gamma_{Lj}$, and $\gamma_{Tj}$ have been
determined using a fit routine developed by A.~Kuzmenko~\cite{kuzmenko}.
Results of these fits are shown for all reflectivity sets of figures~2(a)-2(d).
Despite the use of ceramic samples, the fit works very well. Only close to the
maxima of mode~3, especially for CCO and ZCS, the fit slightly deviates from
the experimental data, probably resulting from a too low density and from
surface imperfections of the ceramic samples.\

Figure~3 shows the central result of this work, namely the temperature
dependence of the eigenfrequencies of four different spinels: geometrically
frustrated CCO, bond-frustrated ZCS, ZCSe, which is an AFM but already
dominated by strong FM exchange, and finally HCS, which is almost
ferromagnetic. The latter two compounds reveal strong metamagnetic behaviour. A
significant temperature dependence of the eigenfrequencies and a splitting of a
number of modes below the magnetic ordering temperatures are clearly
observable. Preliminary results on ZCS and ZCSe were published
in~\cite{hemberger2002, rudolf}. The phonon modes of ZCO have been investigated
in detail in~\cite{sushkov}. CCO, ZCS and ZCSe reveal very different splittings
of the phonon modes in the antiferromagnetic phase.\

\begin{figure}
\begin{center}
\includegraphics[width=11cm]{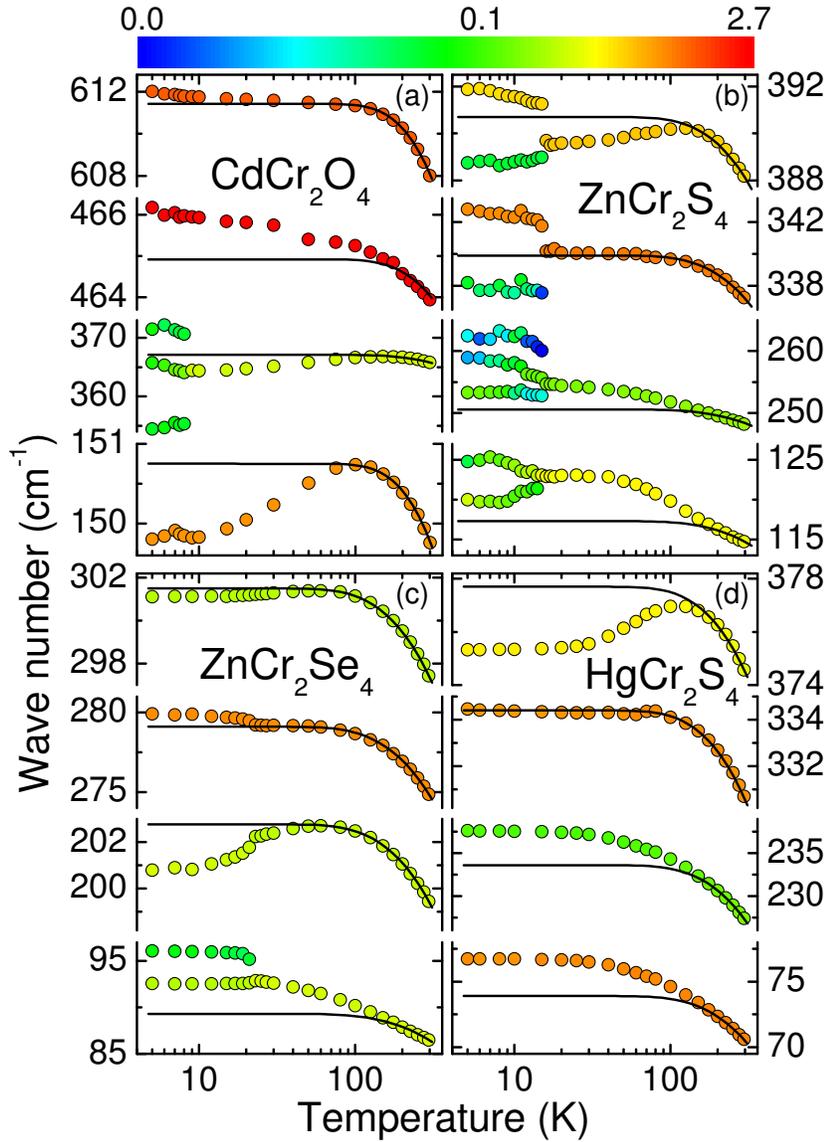}
\end{center}
\caption{\label{fig3}Temperature dependence of the phonon eigenfrequencies for
CdCr$_2$O$_4$(a), ZnCr$_2$S$_4$(b), ZnCr$_2$Se$_4$(c), and HgCr$_2$S$_4$(d).
The mode strength is indicated by a color scale extending from 0.001 (dark
blue) to 2.7 (bright red).}
\end{figure}

CCO exhibits a clear splitting of mode~2 and there are also indications for a
splitting of mode~3. To exemplify these splittings of the phonon modes at the
antiferromagnetic transition of CdCr$_2$O$_4$, figure~4 shows the splitting of
mode~2 and 3 in more detail on an enlarged scale for temperatures just above
and below the magnetic ordering temperature $T_\textrm{N} = 8$~K. An additional
excitation with rather low intensity appears on the low-frequency side of
mode~2. A closer inspection reveals that a third mode is hidden below the main
resonance. This can best be seen by the slight differences of the measured
reflectivity close to 370~cm$^{-1}$, at 9~K and at 5~K. This evolution of a
slight shoulder just below $T_\textrm{N}$ could easily be analyzed with the
used fit routine and is indicated in figure~3. The overall splitting of mode~2
at low temperatures approximately amounts 16~cm$^{-1}$ (see frame (a) of
figure~3). A splitting of mode~2, but only into two modes, has also been
observed by Sushkov {\it et al.}~\cite{sushkov} in ZCO. In this compound a
shoulder appears on the high-frequency side of the main intensity with an
overall splitting of the order of 11~cm$^{-1}$. As documented in figure~4, in
the magnetically ordered phase of CCO a small anomaly also becomes visible
close to 500~cm$^{-1}$. We were not able to fit this apparent splitting by the
automated fit routine and hence, did not include this observation into
figure~3. But we were able to model the experimental results by assuming a
splitting of the mode by approximately 20~cm$^{-1}$. As a splitting of all
modes has been theoretically predicted by Fennie and Raabe~\cite{fennie},
within a first principle model, as well as within LSDA + U, we checked also the
other modes. We found no indications of a splitting, even not in the width
parameter, for mode~4. Mode~1 reveals a slight increase in the width of the
phonon excitation when entering the antiferromagnetic phase, at least some hint
for a possible hidden splitting below experimental resolution. Finally we made
a closer inspection of the phonon modes of ZnCr$_2$O$_4$ at low temperatures.
Modes~2 and 3 are shown in figure~5 on an enlarged scale for clarity. The
splitting of mode~2 is fully compatible with the results of Sushkov {\it et
al.}~\cite{sushkov}. But again we detected a small anomaly at 553~cm$^{-1}$,
which evolves just at the antiferromagnetic phase transition, very similar to
the observations in CdCr$_2$O$_4$. The inset documents that this anomaly is
just on the border of being detectable within our experimental accuracy, but
probably signals a splitting of this mode, similar to the observations in CCO.\
\begin{figure}
\begin{center}
\includegraphics[width=12cm]{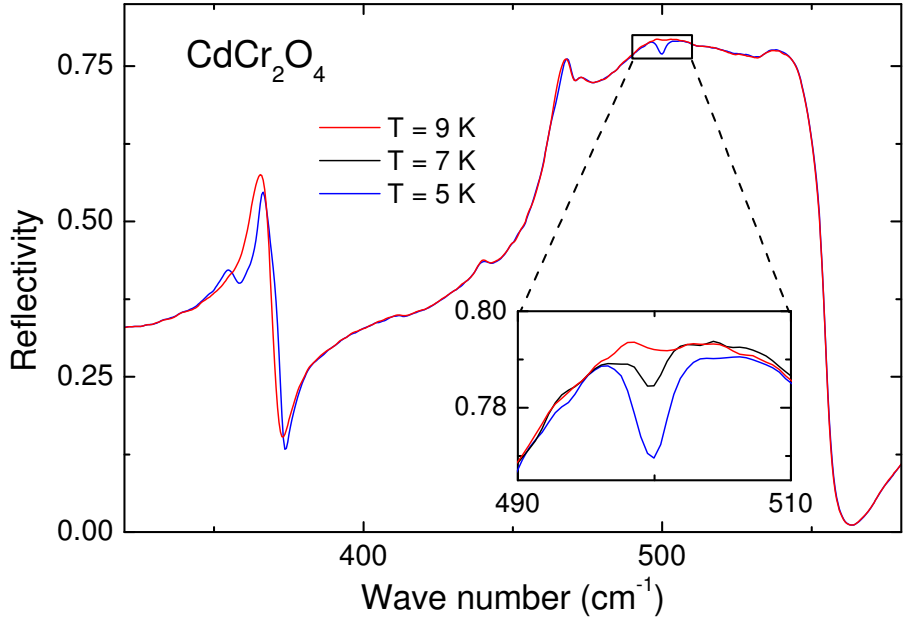}
\end{center}
\caption{\label{fig4}Enlargend scale for the reflectivity of CdCr$_2$O$_4$ for
temperatures around the antiferromagnetic ordering temperature $T_\textrm{N} =
8$~K. A clear splitting of mode~2 and a weak anomaly close to 500 cm$^{-1}$ are
observed. The latter can be modeled assuming a splitting of mode~3 into two
modes of almost equal strength (see text).}
\end{figure}

Returning to figure~3, in ZCSe only the lowest-frequency mode (mode~1) exhibits
a splitting~\cite{rudolf}. In ZCS all modes are split~\cite{hemberger2002}.
Finally, despite the fact that HCS also undergoes antiferromagnetic ordering
below 23~K, no signature of split modes can be observed, and the phonon spectra
in the antiferromagnetically ordered state look similar to those in
ferromagnetic CdCr$_2$S$_4$~\cite{wakamura}.\

In addition to the temperature dependence of the eigenfrequencies, figure~3
also indicates the strength of the modes as function of temperature. To get an
estimate of the dielectric strength, $\Delta\epsilon$ has been calculated via
\begin{equation}
\label{equ3}
\Delta\epsilon_j=\epsilon_{\infty}\left(\frac{\omega_{Lj}^2-\omega_{Tj}^2}{\omega_{Tj}^2}\right)\prod_{i=j+1}\frac{\omega_{Li}^2}{\omega_{Ti}^2},
\end{equation}
which is indicated in figure~3 as colour code, extending over three decades and
running from 0.001 (dark blue for some of the low-intensity split modes) up to
2.7 (dark red, mainly for mode~3). In IR experiments the strength of a mode is
directly related to the ionicity of a bond, or to the effective charges of the
vibrating ions~\cite{wakamura}. Mode~3 is rather strong in all compounds. It
involves displacements of the Cr vs. the $X$ ions forming predominantly ionic
bonds. It is also evident that the IR-active phonons are intense in the oxide
and weak in the selenide, indicating the decreasing electronegativity of the
anions and consequently the increasing tendency for the formation of partly
covalent bonds in the selenide.\

In what follows, we would like to focus on the temperature dependence of the
phonon eigenfrequencies: Starting from high temperatures the frequencies of the
phonon modes increase as usually observed in anharmonic crystals. To get an
estimate of this anharmonic contribution, the temperature dependence was fitted
assuming
\begin{equation}
\label{equ4}
\omega_j=\omega_{0j}\left(1-\frac{c_j}{\textrm{exp}(T/\Theta)-1}\right).
\end{equation}
Here $\omega_{0j}$ indicates the eigenfrequency of mode $j$ at 0~K,
$c_j$ is a mode dependent scale factor of the anharmonic
contributions and $\Theta$ denominates a rough estimate of the Debye
temperature as determined from an average of the four IR active
phonon frequencies. For the analysis the following Debye
temperatures have been determined: CCO: $\Theta=571$~K, ZCS:
$\Theta=392$~K, ZCSe: $\Theta=309$~K, and HCS: $\Theta=361$~K. The
result of these fits to the high-temperature eigenfrequencies ($T
> 100$~K) are indicated for all modes as solid lines in figure~3. It is
clear that all modes reveal significant deviations from this purely anharmonic
behaviour for $T < 100$~K. The deviations are of the order of some percent and
are positive for mode~1 and mode~3, but negative for modes~2 and 4. In addition
to these anomalies in the temperature dependence, some modes reveal a clear
splitting below the corresponding antiferromagnetic phase transitions.\
\begin{figure}
\begin{center}
\includegraphics[width=12cm]{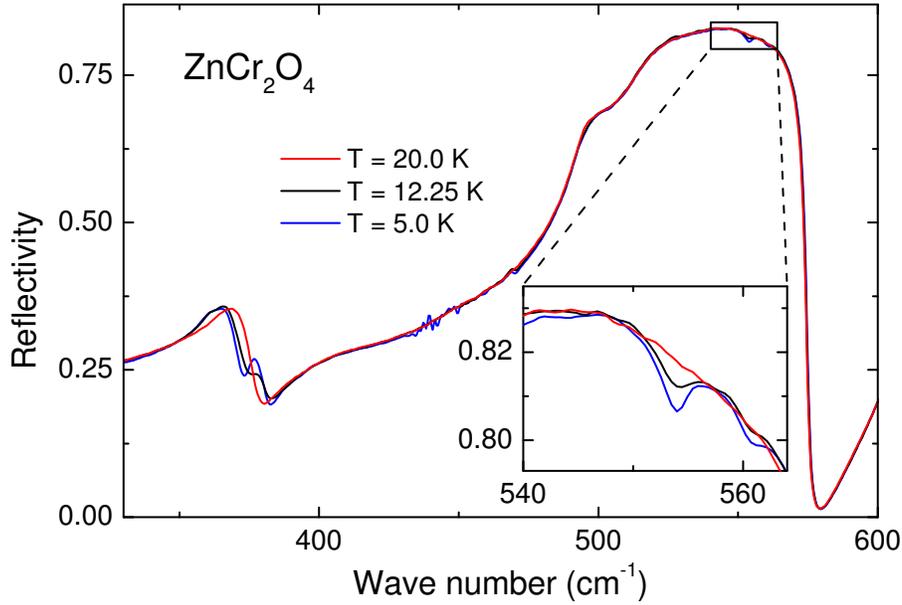}
\end{center}
\caption{\label{fig5}Partial view of the reflectivity spectra of
ZnCr$_2$O$_4$ for temperatures close to the N\'{e}el temperature of
$T_\textrm{N} = 12.5$~K. In addition to the clear splitting of
mode~2 also mode~3 shows a temperature dependent anomaly at about
555~cm$^{-1}$, comparable to CdCr$_2$O$_4$.}
\end{figure}

The pattern of these splittings is very different for the different compounds.
This has been outlined in detail by Tchernyshyov {\it et
al.}~\cite{tchernyshyov, tchernyshyov66} showing that the details of the
particular N\'{e}el order will depend on the details of lattice distortions and
bond order established by the spin-Teller effect. CCO is a geometrically
frustrated magnet and undergoes antiferromagnetic order into a spiral structure
at 6~K. Here mode~2 and 3 split in the AFM phase. Mode~2 splits into three
branches, while for mode~3 a small anomaly evolves at low temperatures, which
hardly can be analyzed (see figure~4). Figure~3 clearly documents that in ZCS,
which is bond-frustrated~\cite{lee} and reveals a complex collinear spin
structure in the ground state, all phonon modes split. Finally, ZCSe and HCS,
which undergo AFM phase transitions into spiral spin phases and are dominated
by ferromagnetic exchange, again behave different: In ZCSe only mode~1 splits,
a splitting which can be fully suppressed in an external magnetic field of
6~T~\cite{rudolf}. HCS shows no splitting at all. Nevertheless, the
eigenfrequencies reveal significant shifts when entering the AFM phases.\

\begin{figure}
\begin{center}
\includegraphics[width=11cm]{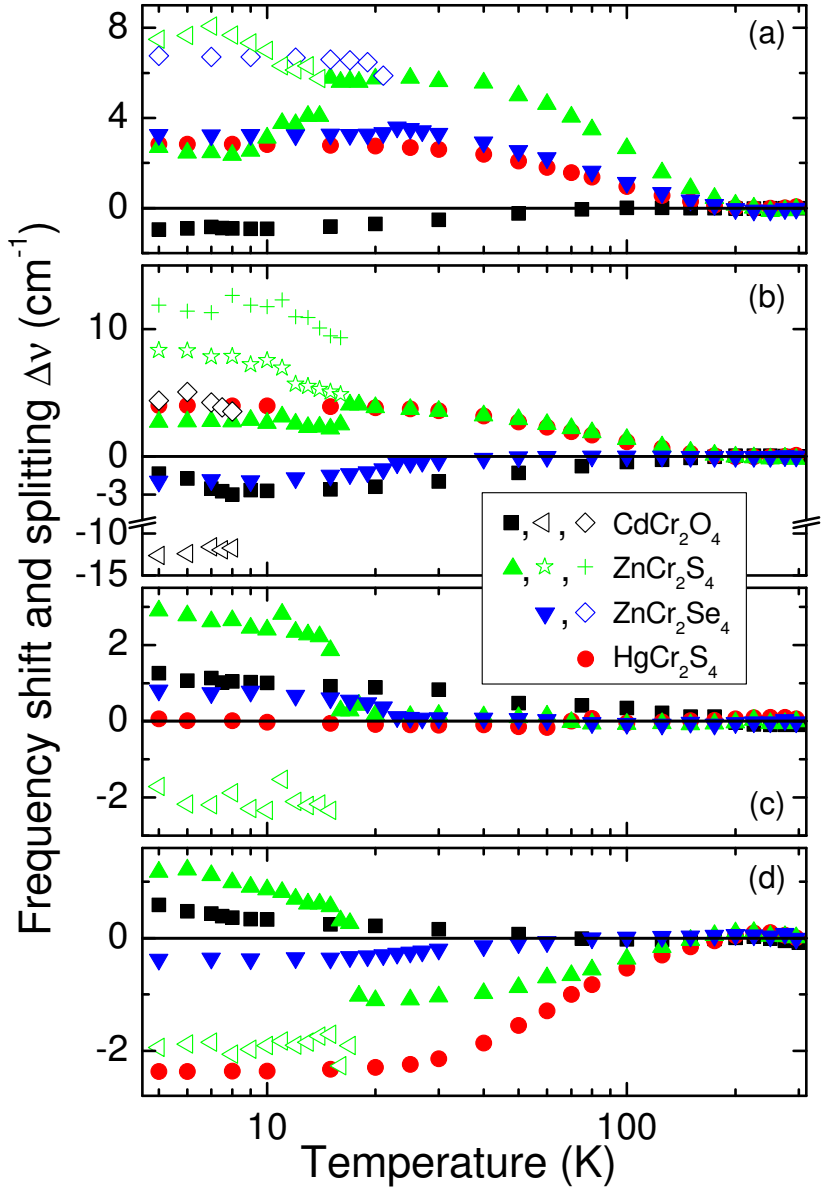}
\end{center}
\caption{\label{fig6}Temperature dependence of shift and splitting of all
eigenfrequencies for CdCr$_2$O$_4$ (black symbols), ZnCr$_2$S$_4$ (green
symbols), ZnCr$_2$Se$_4$ (blue symbols), and HgCr$_2$S$_4$ (red circles) for
modes~1(a), 2(b), 3(c), and for mode~4(d).}
\end{figure}
To analyse these temperature-induced shifts of the eigenfrequencies in more
detail, we calculated the differences between measured frequencies and the
calculated and extrapolated anharmonic behaviour (solid lines in figure~3).
These differences for all modes are shown in figure~6. For mode~1 (figure~6(a))
the shifts in the almost ferromagnetic sulfides and selenides are positive and
amount approximately 5~cm$^{-1}$. It is small and negative for geometrically
frustrated CCO. A splitting can only be observed for ZCS and ZCSe and amounts
5~cm$^{-1}$ and 3~cm$^{-1}$, respectively. For mode~2 (figure 6(b)), ZCS and
HCS exhibit positive, CCO and ZCSe negative shifts. With the exception of
almost ferromagnetic HCS and ZCSe, the phonons for all other compounds split
below $T_\textrm{N}$. For mode~3 (figure 6(c)) the shift for all compounds is
small and positive. ZCS and CCO (not analyzed) exhibit a splitting. Finally,
mode~4 in HCS, ZCS, and ZCSe shows negative shifts, while CCO has a small
positive shift (figure 6(d)). Only ZCS exhibits a splitting of mode~4. It is
interesting to note, that specifically for HCS and ZCS a strong influence of
spin-phonon coupling on the eigenfrequencies can already be detected at 100~K,
obviously determined by the strong FM spin fluctuations (see figure~1).\

The results of figure~6 can be directly compared with theoretical
models: The frequency shift corresponds to the spin-correlation
function times a spin-phonon coupling constant
$\lambda$~\cite{sushkov, fennie, baltensperger1968, lockwood,
wesselinowa}:\
\begin{equation}
\label{equ5} \omega=\omega^0+\lambda<S_i \cdot S_j>
\end{equation}

Here $\omega$ corresponds to a renormalized phonon frequency, while $\omega^0$
is the eigenfrequency in the absence of spin-phonon coupling. $<S_i \cdot S_j>$
denotes a statistical average for adjacent spins~\cite{lockwood}.\

Assuming typical values for EuO, Baltensperger and
Helman~\cite{baltensperger1968} arrived at an estimate of -0.2\% for an average
frequency shift. From figure~6 we find that the shifts are of the order of some
cm$^{-1}$ for most compounds, yielding relative shifts of some per cent for the
low-frequency modes and of less than 1\% for the high-frequency modes. The
splitting has been calculated for ZnCr$_2$O$_4$ by Fennie and Rabe from first
principle methods. These authors find splittings of the order of 5-10~cm$^{-1}$
for the low-frequency modes and of 2-3~cm$^{-1}$ for the high frequency modes,
which reasonably compare to our results.\

Wakamura and Arai calculated the frequency shifts below $T_c$ in ferromagnetic
CdCr$_2$S$_4$ and assumed negative contributions to the mode frequency from
ferromagnetic exchange and positive frequency shifts arising from
antiferromagnetic exchange. Below $T_c$, the authors experimentally found
positive shifts for the low-frequency modes~1 and 2 and negative shifts for the
high frequency modes~3 and 4. They concluded that the ferromagnetic Cr-S-Cr
bonds dominate modes~3 and 4, while the antiferromagnetic linkage Cr-S-Cd-S-Cr
determines the positive shifts of the eigenfrequencies of modes~1 and 2. Our
experimental observations in almost ferromagnetic HgCr$_2$S$_4$ are very
similar, with large posive shifts for modes~1 and 2 and a negative shift for
mode~4. The eigenfrequencies of mode~3 remain almost unshifted and reveal the
normal anharmonic temperature dependence. These frequency shifts due to the
spin-phonon coupling set in already at high temperatures, documenting that HCS
is dominated by ferromagnetic fluctuations. The onset of antiferromagnetic
order at 22~K gives no fingerprint in the temperature dependence of the
eigenfrequencies. This pattern of frequency shifts is similar for most
compounds shown in figure~6. HCS, ZCSe and ZCS exhibit a spiral spin order at
low temperatures, which in the case of ZCS is followed by a mixed phase of
spiral order coexisting with a complex collinear order. Only antiferromagnetic
geometrically frustrated CCO behaves different with positive shifts of modes~3
and 4 and negative shifts of modes~1 and 2. And indeed, the magnetic order of
CCO can best be described as complex collinear state~\cite{chern, chung}.

\section{Conclusions}

The magnetic ions in chromium spinels (spin-only $S = 3/2$, no spin-orbit
coupling) reside on a pyrochlore lattice, which in the case of next-nearest
antiferromagnetic exchange is strongly frustrated. But in addition, depending
on the size of the anions, FM and AFM exchange compete. It is this coexistence
of geometrical frustration and bond frustration which constitutes a variety of
complex ground states as function of Cr-Cr separation or, as we have plotted in
figure~1, as function of CW temperatures. The frustration in all compounds,
geometric as well as bond frustration, seems to be released by a spin-driven
Jahn-Teller effect~\cite{yamashita, tchernyshyov, rudolf}, inducing lattice
anomalies and strong symmetry breaking in the dynamic variables.\

In this report we presented a detailed study of the phonon properties in the
paramagnetic and antiferromagnetic states. We observed strong shifts of the
eigenfrequencies and characteristic splittings of the phonon modes, which are
different for the different order of the magnetic moments. Frequency shifts and
mode splitting are compared for all compounds and are related to a
spin-correlation function of nearest neighbour spins. Moreover, the temperature
dependence of the splittings and shifts seem to follow the sublattice
magnetisation in the ordered compounds. This seems to be best documented for
ZnCr$_2$S$_4$, which shows a splitting of all modes below $T_\textrm{N}$ (green
symbols in panels (a)-(d)). But for some modes, short-range order effects
extending to temperatures far above $T_\textrm{N}$ are clearly observed.
Positive and negative shifts can be related to dominating FM and AFM exchange
acting on specific bonds, which are actively involved in the IR-active modes.\

Finally, we also analyzed the intensity of the different modes and
found, as expected from simple electronegativity considerations,
decreasing mode intensities when moving from the oxide over the
sulphide to the selenides. This decreasing mode intensities signal
an increasing covalency of the bonds, with CdCr$_2$O$_4$ being
strongly ionic, while ZnCr$_2$Se$_4$ being predominantly covalent
bonded.\

\section{Acknowledgments}
This work was supported by the Collaborative Research Center SFB~484
(Augsburg). The support of US CRDF and MRDA via BGP III grant
MP2-3050 is also gratefully acknowledged.


\section{References}
\begin{thebibliography}{30}

\bibitem{baltzer}
Baltzer P K, Wojtowicz P J, Robbins M and Lopatin E 1966, {\it
\PR}{\bf 151} 367--77
\bibitem{hemberger2002}
Hemberger J, Rudolf T, Krug von Nidda H-A, Mayr F, Pimenov A,
Tsurkan V and Loidl A 2006, {\it \PRL}{\bf 97} 087204
\bibitem{lee}
Lee S-H, Broholm C, Ratcliff W, Gasparovic G, Huang Q, Kim T H and
Cheong S-W 2002, {\it Nature} {\bf 418} 856--8
\bibitem{hemberger2006}
Hemberger J, Krug von Nidda H-A, Tsurkan V and Loidl A, Colossal
magnetostriction and negative thermal expansion in the frustrated
antiferromagnet ZnCr$_2$Se$_4$ {\it Preprint} cond-mat/0607811
\bibitem{tsurkan}
Tsurkan V, Hemberger J, Krimmel A, Krug von Nidda H-A, Lunkenheimer
P, Weber S, Zestrea V and Loidl A 2006, {\it \PR}B {\bf 73} 224442
\bibitem{hemberger2005}
Hemberger J, Lunkenheimer P, Fichtl R, Krug von Nidda H-A, Tsurkan V
and Loidl A 2005, {\it Nature} {\bf 434} 364--7
\bibitem{weber}
Weber S, Lunkenheimer P, Fichtl R, Hemberger J, Tsurkan V and Loidl
A 2006, {\it \PRL}{\bf 96} 157202
\bibitem{kino}
Kino Y and L\"{u}thi B 1971, {\it \SSC}{\bf 9} 805--8
\bibitem{yamashita}
Yamashita Y and Ueda K 2000, {\it \PRL}{\bf 85} 4960--3
\bibitem{tchernyshyov}
Tchernyshyov O, Moessner R and Sondhi S L 2002, {\it \PRL}{\bf 88}
067203
\bibitem{tchernyshyov66}
Tchernyshyov O, Moessner R and Sondhi S L 2002, {\it \PR}B {\bf 66}
064403
\bibitem{rudolf}
Rudolf T, Kant Ch, Mayr F, Hemberger J, Tsurkan V and Loidl A,
Spin-phonon coupling in ZnCr$_2$Se$_4$ {\it Preprint}
cond-mat/0611041
\bibitem{sushkov}
Sushkov A B, Tchernyshyov O, Ratcliff II W, Cheong S W and Drew H D
2005, {\it \PRL}{\bf 94} 137202
\bibitem{massida}
Massidda S, Posternak M, Baldereschi A and Resta R 1999, {\it \PRL}{\bf 82}
430--3
\bibitem{fennie}
Fennie C J and Rabe K M 2006, {\it \PRL}{\bf 96} 205505
\bibitem{baltensperger1968}
Baltensperger W and Helman J S 1968, {\it Helv. Phys. Acta} {\bf 41}
668--73
\bibitem{baltensperger1970}
Baltensperger W 1970, {\it \JAP}{\bf 41} 1052--4
\bibitem{bruesch}
Br\"{u}esch P and D'Ambrogio F 1972, {\it \PSS}(B) {\bf 50} 513--26
\bibitem{wessels}
Wessels A L, Czekalla R and Jeitschko W 1997, {\it Mater. Res. Bull.} {\bf 33}
95--101
\bibitem{ueda} Ueda H, Mitamura H, Goto T and
Ueda Y 2006, {\it \PR}B {\bf 73} 094415
\bibitem{chern}
Chern G-W, Fennie C J and Tchernyshyov O 2006, {\it \PR}B {\bf 74}
060405(R)
\bibitem{chung}
Chung J-H, Matsuda M, Lee S-H, Kakurai K, Ueda H, Sato T J, Takagi
H, Hong K-P and Park S 2005, {\it \PRL}{\bf 95} 247204
\bibitem{hamedoun}
Hamedoun M, Wiedenmann A, Dormann J L, Nogues M and Rossat-Mignod J
1986, {\it \JPC}{\bf 19} 1783--800
\bibitem{plumier1965}
Plumier R 1965, {\it Compt. Rend. Acad. Sci. Paris} {\bf 260} 3348--50
\bibitem{plumier1966}
Plumier R 1966, {\it J. Physique} {\bf 27} 213--9
\bibitem{akimitsu}
Akimitsu J, Siratori K, Shirane G, Iizumi M and Watanabe T 1978,
{\it \JPSJ}{\bf 44} 172--80
\bibitem{chapon}
Chapon L C, Radaelli P G, Hor Y S, Telling M T F and Mitchell J F,
Non-collinear long-range magnetic ordering in HgCr$_2$S$_4$ {\it
Preprint} cond-mat/0608031
\bibitem{dwight}
Dwight K and Menyuk N 1967, {\it \PR}{\bf 163} 435--43
\bibitem{lutz}
Lutz H D, W\"{a}schenbach G, Kliche G and Haeuseler 1983, {\it J. Solid
State Chem.} {\bf 48} 196--208
\bibitem{zwinscher}
Zwinscher J and Lutz H D 1995, {\it J. Solid State Chem.} {\bf 118}
43--52
\bibitem{kuzmenko}
RefFIT by A. Kuzmenko, University of Geneva, Version 1.2.44 (2006),
\url{http://optics.unige.ch/alexey/reffit.html}
\bibitem{wakamura}
Wakamura K and Arai T 1988, {\it \JAP}{\bf 63} 5824--9
\bibitem{lockwood}
Lockwood D J and Cottam M G 1988, {\it \JAP}{\bf 64} 5876--8
\bibitem{wesselinowa}
Wesselinowa J M and Apostolov A T 1996, {\it \JPCM}{\bf 8}, 473--88

\end{thebibliography}
\end{document}